# Assessing the Viability of Quantum-Resistant IKEv2 over Constrained and Internet-Scale Networks

Geoff Twardokus
ESL Global Cybersecurity Institute
Rochester Institute of Technology
Rochester, NY, USA
gdt5762@rit.edu

William Joslin
Rochester Institute of Technology
Rochester, NY, USA
wpj3799@rit.edu

Hanif Rahbari
ESL Global Cybersecurity Institute
Rochester Institute of Technology
Rochester, NY, USA
hxrics@rit.edu

William Layton
National Security Agency
Washington, D.C., USA
William.Layton@cyber.nsa.gov

## Abstract

Within 1–2 decades, quantum computers may become powerful enough to break current public-key cryptography, prompting authorities such as the IETF and NIST to push for adopting quantum-resistant cryptography (QRC) in ecosystems like Internet Protocol Security (IPsec). Yet, IPsec struggles to adopt QRC, primarily because Internet Key Exchange Protocol Version 2 (IKEv2), which sets up IPsec sessions, cannot easily tolerate the large public keys and digital signatures of QRC. Many IETF RFCs have been proposed to integrate QRC into IKEv2, but their performance and interplay remain largely untested in practice. In this paper, we measure the performance of these RFCs over constrained links by developing a flexible, reproducible measurement testbed for IPsec with quantum-resistant IKEv2 proposals. Deploying our testbed in lossy wireless links and on the internationally distributed FABRIC testbed for Internet scenarios, we reveal that bottlenecks arise with quantum-resistant IKEv2 under high round-trip times, non-trivial packet loss, or other constraints. Our results, including the revelation of a 400–1000-fold increase in data overhead over high-loss wireless links, expose the shortcomings of today's RFCs and call for further work in this vital area of post-quantum network security.

**ACM Reference Format:**
Geoff Twardokus, William Joslin, Hanif Rahbari, and William Layton. 2025. Assessing the Viability of Quantum-Resistant IKEv2 over Constrained and Internet-Scale Networks. In *Proceedings of the 2025 Quantum Security and Privacy Workshop (QSec '25), October 13–17, 2025, Taipei, Taiwan.* ACM, New York, NY, USA, 6 pages. https://doi.org/10.1145/3733825.3765281

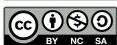

## 1 Introduction

Internet Protocol (IP) networks are indispensable in modern society, with essential activities such as e-commerce, remote work, and virtual healthcare relying on fast and reliable IP communication. IP also plays a vital role in national security, as government operations, ranging from real-time diplomacy to next-generation control systems for nuclear weapons [28], depend heavily on public and classified IP networks. These sensitive communications occur over potentially constrained air-, land-, and space-based links that are vulnerable to interception, tampering, and other attacks; therefore, the ability to communicate confidentially with *authenticated* parties is essential. Today, this ability is provided primarily through the IP Security (IPsec) protocol suite [13], utilizing cryptographic algorithms approved by the National Institute of Standards and Technology (NIST). IPsec today features lightweight secure session setup, which is critical for time-sensitive or real-time operations over possibly constrained links.

In fact, IPsec connection setup today requires the exchange of just four messages using Internet Key Exchange Protocol Version 2 (IKEv2) [14]. Employing fast and lightweight cryptographic algorithms such as Diffie-Hellman Key Exchange (DHKE) and the Elliptic Curve Digital Signature Algorithm (ECDSA), IKEv2 has been designed and tested for efficiency [24]. However, the accelerating development of quantum computers imperils the future of IKEv2, and by extension, IPsec, as such computers will eventually be able to break these public-key algorithms. Anticipating the quantum threat, NIST has standardized new *quantum-resistant* cryptographic algorithms, or QRC [25], to replace such classical algorithms. Underscoring the urgency, U.S. legislation and executive orders [26, 27] as well as E.U. regulations [11] require their respective governments to immediately prepare for and undertake a transition to QRC. Notably, the National Security Agency (NSA) in the U.S. now requires IPsec in national security systems to transition to QRC by 2030 [23]. To this end, the Internet Engineering Task Force (IETF) is considering several proposed standards—known as Requests for Comments, or RFCs—towards quantum-resistant IPsec.

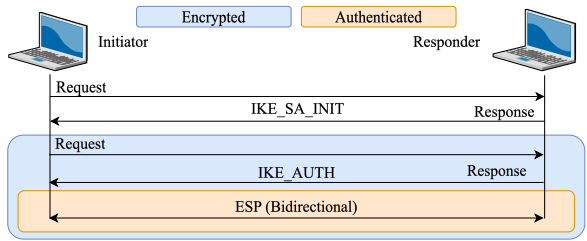

**Figure 1: Classical IKEv2 IPsec connection setup.**

However, this momentum is challenged by the practical reality that IPsec struggles to accommodate the much larger signatures and keys of QRC, which significantly exceed the payload size limit of Layer 2 protocols such as Ethernet. Ordinarily, IP-layer fragmentation would offer a straightforward solution; however, since intermediary network devices and firewalls are often configured to drop fragmented IP packets [15], this approach could make connection establishment difficult over uncontrolled networks like the Internet. The fragmentation problem (which goes beyond QRC) drove the development of IKEv2-specific fragmentation in RFC 7383 [15], which underpins several recent RFCs for quantum-resistant IKEv2 (e.g., [17, 19]). However, these RFCs were not designed with constrained networks in mind. More packets, longer QRC keys, and the interplay between different RFCs may even introduce new problems—for example, as we will show, by dramatically increasing the number of packets that must be (re-)transmitted to connect over a lossy network.

In this paper, we design a measurement strategy to shed light on obstacles, identifying real-world bottlenecks that proposed RFCs (e.g., [15–19]) would create in today's Internet and potential opportunities to guide future efforts. We show such approaches have serious performance issues in networks with higher round-trip times (RTTs), non-trivial packet loss, or other constraints. These findings have been overlooked until now, as no prior work has meaningfully measured IKEv2 performance with QRC beyond small-scale, isolated scenarios (e.g., [6, 8]) that do not represent practical environments with bandwidth constraints, lossiness, or other limitations.

To this end, we develop the first comprehensive and reproducible measurement framework for quantum-resistant IKEv2 that facilitates performance studies in large-scale distributed Internet testbeds, wireless networks with commercial hardware, and other realistic scenarios. Using this open-source testbed, we expose serious shortcomings in current RFCs for quantum-resistant IKEv2—most notably, that on severely impaired wireless links, using QRC can require transmitting over 1000 times as much data in IKEv2 as with classical cryptography. Our specific contributions are as follows:

- We curate state-of-the-art, interconnected RFC proposals for quantum-resistant IKEv2 (e.g., [15–19]) and scrutinize their challenges in constrained networks, showing that these existing proposals are not yet sufficient to accommodate QRC in IKEv2.
- We engineer *measuredSwan*, a flexible and open-source testbed for quantum-resistant IKEv2 that supports a range of customizable, real-world experiments. *measuredSwan*[1] extends the strongSwan IPsec suite [2] with measurement scripts and timing hooks[2] embedded within specific RFC functionalities, and augments this core with an automated deployment framework enabling real-world experiments with commodity hardware (e.g., Wi-Fi access points) and Internet measurement on the distributed FABRIC testbed [5].
- We expose significant, potentially catastrophic problems when quantum-resistant IKEv2 is deployed across various realistic scenarios. Our results reveal massive overhead increases—more than 1000-fold—when setting up connections using QRC. We identify these as unforeseen consequences of jointly using specific RFCs in networks where packet loss exceeds ∼2%, leading us to deduce that today's RFCs are not yet sufficient for practical deployment of quantum-resistant IKEv2 in wireless and other constrained scenarios.

The remainder of this paper is structured as follows. In Section 2, we provide technical context for quantum-resistant IKEv2. We present our testbed and measurement results in Sections 3 and 4, respectively, and conclude with Section 5.

## 2 IPsec in Transition

IPsec is defined across several RFCs centered on [13], but it broadly breaks down into two protocols.

First, connection setup, including key establishment and endpoint authentication, is accomplished using the four-message process shown in Figure 1. Keys are established using DHKE via the initialization messages (IKE_SA_INIT), allowing subsequent messages to be encrypted, while authentication is completed using certificates, digitally signed with ECDSA or Rivest–Shamir–Adleman (RSA), that are exchanged in the IKE_AUTH messages. Problematically, DHKE, ECDSA, RSA, etc., are all susceptible to quantum attacks like Shor's Algorithm, making IKEv2 a potential Achilles' heel for IPsec in the post-quantum era.

Successful IKEv2 setup produces an authenticated and encrypted *security association*, or SA, within which secure data communication can commence using Encapsulating Security Payload (ESP), the second protocol that comprises IPsec. The encryption used by ESP (and IKE_AUTH) is nearly always the Advanced Encryption Standard (AES), usually with 256-bit keys (AES-256). As AES-256 retains 128-bit security against the best known quantum attacks [29], ESP is of far less concern than IKEv2 when it comes to quantum resistance.

### 2.1 Proposals for Quantum-Resistant IKEv2

A number of proposals have been put forward to tackle the challenges of adopting QRC in IKEv2. Here, we briefly review the four most accepted and relevant proposals [15, 17, 19, 21], which we use in our testbed. We further discuss limitations of these RFCs that cast doubt on their viability in constrained scenarios. Keeping with our focus on preparing for a near-term transition to QRC in IKEv2, we limit our discussion to proven and/or standardized QRC techniques, such as lattice-based algorithms (see below), and we do not discuss nascent technologies like quantum key distribution (QKD) that are far from practical deployment. We also briefly remark on less mainstream ideas such as [16, 18, 22].

---

[1]https://github.com/rit-wisp-lab/measuredSwan

[2]https://github.com/rit-wisp-lab/pq-strongswan

**IKEv2 Message Fragmentation.** RFC 7383 [15] attempts to resolve the problem of sending IKEv2 messages that exceed the Maximum Transmission Unit (MTU) of Layer 2 by fragmenting data at the *application* layer. This approach obviates the need to fragment packets at the IP-layer, but it comes with a major caveat—only *encrypted* IKEv2 messages can be fragmented this way. Thus, while large IKE_AUTH messages can be fragmented, this approach does not work for the unencrypted IKE_SA_INIT messages used to establish the secure channel (see Figure 1). IP-layer fragmentation of large QRC IKE_SA_INIT messages thus remains unresolved.

**Intermediate Key Exchanges.** RFC 9242 [17] tackles the issues in RFC 7383 by introducing IKE_INTERMEDIATE messages, which use the above IKEv2 fragmentation mechanism to exchange quantum-resistant keys after IKE_SA_INIT and before the IKE_AUTH exchange that completes the SA. An additional notification flag in the initial IKE_SA_INIT message, triggers a cascade of small supplemental messages containing fragments of the public key; thereby avoiding Layer 3 fragmentation. However, this requires exchanging many more ($\geq$ 40 vs. 4) and much larger IKEv2 messages than with classical cryptography—potentially, as we will show, transmitting tens of thousands of more bytes with QRC.

**Multiple Key Exchanges.** RFC 9370 [19] extends RFC 9242 to enable multiple key exchanges during the IKEv2 setup process. This method introduces IKE_FOLLOWUP_KEY messages and the ADDITIONAL_KEY_EXCHANGE notification to facilitate, e.g., the use of hybrid classical/quantum-resistant configurations where at least two keys need to be shared between peers. While an essential part of complying with mandates, e.g., the NSA requirement to move IPsec to hybrid security by 2030 [23], multiple key exchanges only exacerbate the problem of increased message quantity and size introduced by intermediate key exchanges in RFC 9242.

**Quantum-resistant KEMs.** To eliminate quantum-vulnerable DHKE, it has recently been proposed to employ Module Lattice-based Key Encapsulation Mechanism (ML-KEM), to securely establish shared encryption keys during the IKE_SA_INIT phase. One draft builds on Multiple Key Exchange, RFC 9370 [21], which we evaluate. And another draft builds on Announced Supported Authentication Methods, RFC 9593 [20], which we don't consider due to the RFC's unnecessary IKE_INTERMEDIATE overhead. However, even at security level 1, it still requires very large QRC keys that bring us back to the fragmentation problem discussed above.

**Additional Relevant RFCs.** As an alternative to IKEv2 fragmentation, it has been proposed to move (part of) the IPsec protocol from UDP to TCP at the transport layer and avoid fragmentation altogether [18, 22]. However, since applications using the secure ESP data flow would then have to establish their own TCP sessions within an outer IPsec TCP connection, such proposals would create "TCP-in-TCP" sessions that are prone to connection meltdowns. TCP-based solutions have not been widely accepted in the community, so we do not consider them in our evaluations.

Another effort to avoid exchanging large QRC keys (and thus fragmentation), is using pre-shared keys in RFC 8784 [16]. Here, pre-shared quantum-resistant key material is "mixed" into the classical IKEv2 connection setup. This approach is effective in adding quantum-resistant security, but at the non-trivial cost of requiring endpoints to be pre-configured and periodically updated with fresh shared secrets. This overhead is impractical for systems where endpoints change, distances are long, or keys require frequent updates, and so we do not consider this approach in this work.

## 3 A Flexible Testbed for Quantum-Resistant IKEv2

We now present our *measuredSwan* framework for evaluating quantum-resistant IKEv2. *measuredSwan* is built around a flexible, automated deployment architecture for reproducible real-world experiments.

**Software:** The core of our testbed is *measuredSwan*, which we develop as an extension of the popular strongSwan application for setting up IPsec-based virtual private networks (VPNs) [2]. In its latest version 6.0.2, strongSwan natively supports using ML-KEM in DHKE for the initial exchange of IKE_SA_INIT messages, as suggested by [20, 21]. The three "mainstream" proposals for enabling quantum resistance—IKEv2 fragmentation, intermediate key exchanges, and multiple key exchanges (see Section 2)—are also supported. Employing these alongside the liboqs plugin [1], messages can be signed using QRC such as the Module-Lattice-Based Digital Signature Algorithm (ML-DSA) instead of classical ECDSA. In our experiments, we compare classical configuration of IKEv2 with a QRC configuration, as detailed in Table 1. Our evaluations are conducted on hosts running Ubuntu 22.04 LTS.

**Hardware:** In the simplest configuration, real-world experiments using actual computers and networks can be set up by just installing *measuredSwan* on the host PCs, connecting the desired network topology, and conducting experiments. For more specific scenarios beyond simple wired networks, *measuredSwan* is also compatible with any peripheral devices that operate below the transport layer. For example, wireless networks configured with commercial Wi-Fi NICs and cellular eSIM devices are supported. In some of our experiments, we use commercial Wi-Fi access points and commodity laptops for wireless experiments in an enterprise environment.

**Flexible & Automated Deployment:** To support a wide range of potential experimental scenarios, *measuredSwan* is built around a highly customizable and reconfigurable architecture. This *infrastructure as code* (IAC) design facilitates flexible deployment to local PCs, cloud servers (e.g., OpenStack), research computing clusters, and beyond without modification to our implementation. We provide both the *measuredSwan* source code, including all of the timing hooks and measurement points we added into strongSwan, and an automated deployment capability built using Ansible. Essentially, given the IP address for a Debian-based Linux host, our deployment scripts will handle installing liboqs and *measuredSwan*, executing desired experiments with classical and/or QRC configurations, and compiling the results in a digestible format. For example, our experiments over the Internet are conducted by deploying *measuredSwan* on hosts on the globally distributed FABRIC testbed [5], and several of our measurement results use FABRIC to study how real-world Internet constraints impact IKEv2 performance with QRC.

## 4 Experimental Results

Using *measuredSwan*, we now demonstrate how RFCs for quantum-resistant IKEv2 impact IKEv2 performance (and the broader network) in suboptimal or constrained scenarios.

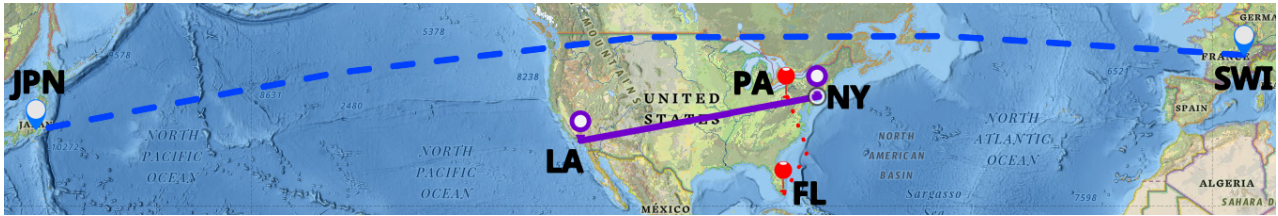

Figure 2: Globally distributed FABRIC links used in our evaluations are between Tokyo (JPN) and Geneva, Switzerland (SWI); Pennsylvania (PA) and Florida (FL); and New York City (NY) and Los Angeles, California (LA).

Table 1: Parameters choices for evaluating IKEv2 with classical vs. quantum-resistant cryptography.

|  | Algorithm | Purpose | Key Length (Bytes) |
|---|---|---|---|
| Classical | AES-256-CBC | Encryption | 32 |
|  | SHA-256 HMAC | Authentication/Integrity | 32 |
|  | DHKE (2048-bit MODP) | Key Establishment | 256 |
| QRC | AES256-CBC | Encryption | 32 |
|  | ML-DSA-87 | Authentication/Integrity | 2592 |
|  | ML-KEM-768 | Key Establishment | 2400 |

## 4.1 Methodology

We consider two distinct testbed environments—one for Internet scenarios, and one for wireless networks.

*4.1.1 Internet Experiments.* For experiments over the Internet, we configure pairs of FABRIC endpoints to act as clients that want to establish a secure IPsec connection over some long distance, as illustrated by Figure 2. The benefits of using FABRIC are twofold. First, IPsec connections between FABRIC endpoints traverse the public Internet and thus capture real-world impairments such as high RTTs, the potential for overwhelmed routers to briefly delay or drop packets, middleboxes with unknown configurations, etc. Second, FABRIC uses a public API that facilitates reproducible experiments using the measurement scripts we provide.

While FABRIC captures many real-world constraints, its deployment on the Internet does not inherently model the lossy links encountered in more constrained networks. Therefore, to study the effects of higher lossiness, we add stochastic packet loss to the endpoints using queuing disciplines ("qdiscs") in the traffic control utility (`tc`) that is built into the Linux kernel. More specifically, we implement the widely used Simplified Gilbert-Elliott (SGE) packet loss model [12]. SGE models bursty packet loss over a link using a two-state Markov model that transitions between a "good" state (no loss) and a "bad" state (100% loss) that simulates, e.g., when a router buffer is overwhelmed and briefly drops all incoming packets. State transitions are based on probabilities $P$ and $R$ of leaving the good or bad state at each step, respectively, yielding the steady-state loss probability:

$$\pi_L = \frac{P}{P+R}. \tag{1}$$

We select values of $P$ and $R$ in our FABRIC experiments such that $\pi_L$ varies between 0–10%, thus covering both typical, more limited loss rates as well as the more severe loss rates that could be encountered in constrained network scenarios.

*4.1.2 Wireless Networks.* To study how IPsec performs in wireless networks where noise, non-line-of-sight channels, and other factors can result in less reliable communication than typical Internet scenarios, we deploy *measuredSwan* on laptop PCs running Ubuntu Linux that communicate with each other via an enterprise 802.11 (Wi-Fi) network on our university campus. By carefully positioning the laptops at such a distance from access points that path loss and fading reduces packet loss to values of interest, we are able to study how integrating QRC into IKEv2 impacts performance in realistically constrained wireless scenarios.

Table 2: Experimental scenarios for FABRIC.

| Site Pair | Mean RTT (ms) | $P$ | $R$ | $\pi_L$ |
|---|---:|---|---|---|
| FL–PA–2 | 31.408 | $3.0 \times 10^{-3}$ | $128.0 \times 10^{-3}$ | 2.29% |
| LA–NY–1 | 66.015 | $4.0 \times 10^{-3}$ | $81.0 \times 10^{-3}$ | 4.71% |
| FL–PA–1 | 31.387 | $2.5 \times 10^{-3}$ | $43.1 \times 10^{-3}$ | 5.3% |
| JPN–SWI | 259.319 | $0.6 \times 10^{-3}$ | $7.7 \times 10^{-3}$ | 7.34% |
| LA–NY–2 | 66.035 | $9.0 \times 10^{-3}$ | $82.0 \times 10^{-3}$ | 9.89% |

*4.1.3 Metrics.* In the literature, the impact of QRC on network security protocols has been studied using metrics such as encryption or key generation time in hardware [8], data rate throttling [10], and connection setup delays [3, 7–9]. Execution time in hardware is not our focus, and data rate is not very meaningful to us because IKEv2 is only used for connection setup, not sustained communication (for which ESP is used instead). Therefore, our first metric is *connection setup time*—i.e., the time that elapses from the first `IKE_SA_INIT` transmission until an SA is successfully established. Intuitively, we expect QRC will increase connection setup time because more packets must be transmitted (tens of packets with QRC vs. 4 with classical), although our results tell a more complex story.

Our second metric, *connection setup data consumption*, is more tailored to focus on IKEv2. We select this metric based on our observation that because IKEv2 is carried over UDP, there is no way to determine which or how many packets were lost in transit if a connection setup attempt fails. Critically, this means that *if even one packet* is lost during the initial connection attempt, the next attempt requires retransmitting *every* IKEv2 packet—not just the one(s) that were lost. We hypothesize this could result in a massive increase in bandwidth consumption by IKEv2 when trying to establish an IPsec connection with QRC over a lossy link.

## 4.2 Connection Setup Time

We begin by measuring how QRC impacts the IKEv2 connection setup time. To do this, we deploy three pairs of FABRIC endpoints with varying physical and network distance (FL-PA-2, LA-NY-1, and JPN-SWI in Table 2). For now, though, we do not add any stochastic

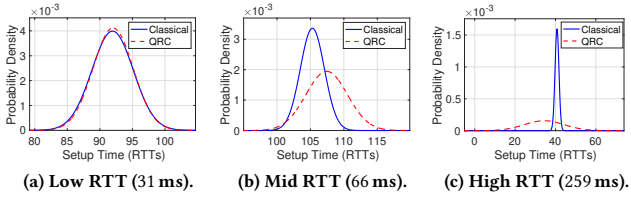

**(a)** Low RTT (31 ms). **(b)** Mid RTT (66 ms). **(c)** High RTT (259 ms).

**Figure 3: IKEv2 connection setup times over 1000 iterations with classical vs. QRC, normalized by RTT.**

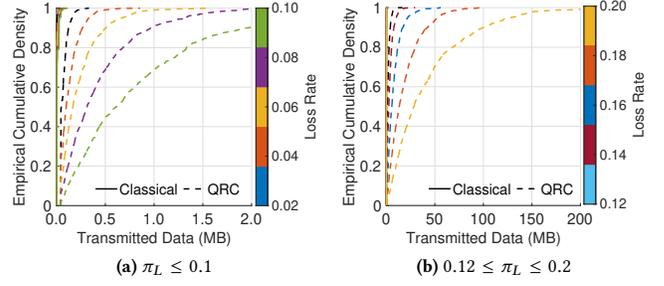

**(a)** $\pi_L \leq 0.1$ **(b)** $0.12 \leq \pi_L \leq 0.2$

**Figure 4: Empirical CDFs for total data transmitted during IKEv2 connection setup with classical vs. QRC at packet loss rates of (a)** 0–10% **and (b)** 10–20%**.**

packet loss (i.e., $\pi_L = 0$), and we consider only whatever loss is naturally present in the Internet routes between endpoints. On each pair of endpoints, we then run 1000 IKEv2 connection setups with the classical and QRC configurations described in Table 1.

Our first observation from the results (Figure 3) is that over reliable links with negligible packet loss, QRC has relatively little impact on connection setup time. Figure 3(a) and 3(b) show that QRC does slightly lengthen connection setup, which is expected due to the transmission of more packets with QRC, although Figure 3(c) counterintuitively shows that setup with QRC sometimes actually takes *less* time than with the classical configuration. We posit this is because Internet routers sometimes prioritize longer UDP streams over shorter bursts, a behavior designed to enhance streaming applications (see, e.g., [4]) that also works in favor of IKEv2. However, irrespective of which configuration runs faster, Figure 3 shows the difference in setup times is on the order of a few dozen milliseconds, which is relatively trivial.

Our second observation is more telling. As Figure 3 shows, it turns out that when the average RTT increases, the distributions of classical and QRC setup times steadily decorrelate (from a coefficient of $r = 0.994$ in Figure 3(a) to just 0.325 in 3(c)). This reflects the greater instability of connection setup when using QRC, which we attribute to the large number of UDP messages transmitted under RFCs 9242 and 9370, and the correspondingly greater chance that intermediary routers may drop or otherwise influence packet flow to improve or degrade connection setup time. Further refinement of proposals for TCP in IKEv2 (see Section 2), or exploring alternatives like QUIC, might help resolve this instability. We encourage more study in this area as a part of future work.

### 4.3 Data Consumption

To measure the amount of data consumed by IKEv2, we consider both FABRIC and wireless scenarios, with FABRIC used to study packet loss rates up to 10% and our wireless topology used to study losses between 10–20% (which are too high for realistic Internet scenarios, but reasonable for mobile or otherwise strongly impaired wireless channels). On FABRIC, we use the configurations and SGE loss model parameters shown in Table 2, while for our wireless scenario we use the configuration described in Section 4.1.2. We run 1000 iterations each of the classical and QRC IKEv2 configurations in each scenario and at each loss rate, this time tracking how much total data is transmitted, by both endpoints, before the connection is successfully established.

Figure 4(a), which shows our FABRIC results as empirical CDFs of each experimental run, is revealing. While classical configurations all remain clustered near the y-axis, QRC runs sharply diverge, indicating substantially more data is transmitted when QRC is used over lossier links. For example, at 8% packet loss, the 95th percentile is 25.6 kB for classical cryptography vs. 1.3 MB for QRC—a 50-fold increase indicating significantly degraded performance with QRC.

The effect is even more pronounced in our wireless results (Figure 4(b)), where at 12% packet loss, the 95th percentile spikes to 22.1 MB with QRC, compared to just 49.8 kB classically—a 443-fold increase. By 20% packet loss, the worst-case scenarios with QRC can result in nearly 200 MB per connection setup, a more than 1000-fold increase from the classical configuration. This level of overhead is likely untenable in any rate-limited or low-bandwidth deployment, showcasing the insufficiency of current RFCs for such networks. We note that the huge increase in data consumption over lossy links is a direct result of IKEv2's dependence on UDP and the significant number of packets that need to be transmitted when using RFCs 9242 and 9370—the same RFCs we identified as troublesome for connection setup time. Ultimately, our results demonstrate that for both connection setup and data consumption, the interplay between these RFCs, coupled with the use of UDP in IKEv2, makes for potentially severe performance collapse in constrained networks, clearly showing there is more work to be done in making quantum-resistant IKEv2 practical for critical constrained scenarios.

## 5 Conclusions and Future Work

In this paper, we developed a novel testbed for quantum-resistant IKEv2 and orchestrated careful measurements of Internet and wireless networks that exposed serious practical challenges. We showed that among proposals for quantum-resistant IKEv2, RFCs 9242 and 9370 are particularly problematic in constrained networks because they require transmitting significantly more packets over potentially lossy links. Our measurements showed this can untenably drain network resources, as IKEv2 consumes significant bandwidth trying to complete its connection setup, with usage increasing by more than 1000-fold in the worst-case scenarios. Cumulatively, our results demonstrate that current RFCs are inadequate for quantum-resistant IKEv2 in constrained networks, and we argue that more work (e.g., refining TCP-based solutions) is essential to protect critical IP-based systems from future quantum threats.

## Acknowledgments

We thank Benjamin Carini for his contributions in our preliminary efforts that led to this work. This research was supported in part by the NSA under Grant Number H98230-21-1-0317. This material is also based on work supported by the National Science Foundation (NSF) under Award No. 2239931. Any opinions, findings, and conclusions or recommendations expressed in this material are those of the author(s) and do not necessarily reflect the views of the NSF.